\documentclass[a4paper]{article}
\usepackage{url}
\usepackage{bm}
\usepackage{multirow}
\usepackage{algorithm}
\usepackage{algorithmic}
\usepackage{graphicx}
\usepackage{amsmath,amssymb,amsthm,amsfonts,latexsym}
\usepackage{pdflscape}
\usepackage{verbatim}

\author{Li ZHANG, Xuejun LIU, Songcan CHEN}
\title{Detecting Differential Expression from RNA-seq Data with Expression Measurement Uncertainty}
\begin{document}
\maketitle
%\setcounter{page}{1}
%\setlength{\baselineskip}{14pt}
 %\footnotesize
College of Computer Science and Technology, Nanjing University of Aeronautics and Astronautics, Nanjing, 210016, China

\begin{abstract}
High-throughput RNA sequencing (RNA-seq) has emerged as a revolutionary and powerful technology for expression profiling.
% RNA-seq expression data consists of read counts of each gene, and
 Most proposed methods for detecting differentially expressed (DE) genes from RNA-seq are based on statistics that compare normalized read counts between conditions.
% Unfortunately, reads are in general too short to be mapped unambiguously to features of interest, such as genes and isoforms. Therefore, many expression estimation methods account for the read mapping ambiguity.
  However, there are few methods considering the expression measurement uncertainty into DE detection. Moreover, most methods are only capable of detecting DE genes, and few methods are available for detecting DE isoforms. In this paper, a Bayesian framework (BDSeq) is proposed to detect DE genes and isoforms with consideration of expression measurement uncertainty. This expression measurement uncertainty provides useful information which can help to improve the performance of DE detection. Three real RAN-seq data sets are used to evaluate the performance of BDSeq and results show that the inclusion of expression measurement uncertainty improves accuracy in detection of DE genes and isoforms. Finally, we develop a GamSeq-BDSeq RNA-seq analysis pipeline to facilitate users, which is freely available at the website \url{http://parnec.nuaa.edu.cn/liux/GSBD/GamSeq-BDSeq.html}.
\end{abstract}

%\Keywords{RNA-seq, Bayesian method, differentially expressed genes/isoforms, expression measurement uncertainty, analysis pipeline}

\section{Introduction}%
In the past decades, DNA microarray is the most important and widely used technology for gene expression measurement, but recently high-throughput RNA sequencing (RNA-seq) has rapidly become a revolutionary and powerful alternative for transcriptome analysis\cite{RNAsurvey1}.
RNA-seq offers several advantages over microarrays, such as an increased dynamic range, a lower background level and a higher throughput. Moreover, RNA-seq enables many applications not achievable by microarrays including discovering unknown transcripts and alternative splicing\cite{RNAsurvey2,RNAsurvey3}.

\begin{figure}[!htbp]
  \centering
  % Requires
  \includegraphics[width=0.6\textwidth]{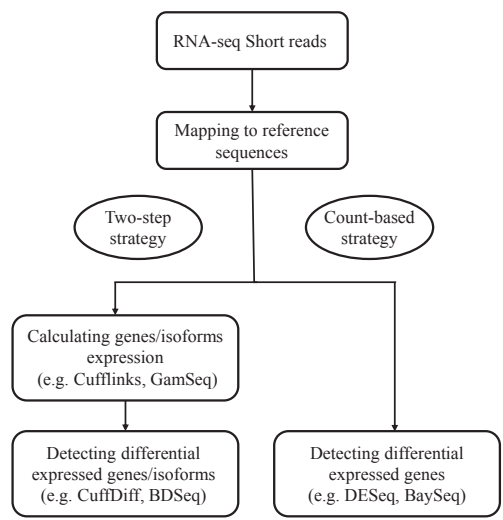}\\
 \caption{The general workflow of differential expression analysis for RNA-seq data.}
 \label{fig0}
\end{figure}

In transcriptome analysis, one fundamental objective is to detect differential expression, i.e., genes and isoforms that show differential expression between conditions\cite{DEsurvey1}.
RNA-seq sequences cDNA fragments that have been derived from an RNA sample and hence produce millions or billions of short subsequence called reads. These reads typically are mapped the a reference genome sequence or the transcriptome sequences.
The number of reads mapped to a genomic feature of interests, which can be a gene, an exon or any region of interest, is used to quantify the abundance of the feature in the analyzed sample\cite{RNAsurvey3}. Subsequently statistical methods are applied to detect the differential expression between conditions\cite{review1}. The general workflow of differential expression analysis is shown in Fig.\ref{fig0}.
Although RNA-seq has several advantages over DNA microarrays,
in reality the analysis of RNA-seq data still remains some difficulties, some of which are inherent to sequencing procedure of RNA-seq experiments, e.g., read mapping ambiguities and sequencing biases. Therefore, detecting differential expression is still an important and key task in transcriptome analysis\cite{RNAsurvey4}.

Recently, many statistical methods are proposed to detect differential expression from RNA-seq data. Most of methods, named count-based methods, are based on statistics that compare read counts between conditions\cite{DEsurvey2}. The read counts are generally normalized by the sequencing depths or library sizes (the total number of mapped reads) for different samples\cite{DEsurvey3}. Most count-based methods commonly use the negative binomial(NB) distribution to address the over-dispersion problem which is caused by the high variability across read counts\cite{review2,review3,DESeq,BaySeq,NBPSeq,sSeq,edgeR,DSS}.
In addition,  Voom estimates the mean-variance relationship and applies the normal linear model to fit read counts\cite{voom}.
NPEBseq developes a novel nonparametric empirical Bayesian-based approach to model RNA-seq data\cite{NPEBSeq}.
These methods are useful for detecting DE genes, because the read counts can be exactly obtained for most genes. When DE isoforms are of interest, researchers often apply the count-based methods directly to estimate isoform counts\cite{countbase}. However, sharing of exons between isoforms of the same genes and sequence homology between genes can result in the read mapping ambiguity. As a result, reads may be aligned to multiple transcripts in the same gene or different genes. Hence, it is difficult to exactly obtain the read counts for each isoform. Therefore, it is not appropriate to use count-based methods for detection of DE isoforms.

Because of read mapping ambiguity, estimating isoform expression is a difficult task. A distinct advantage of RNA-seq is that sequencing along splice junctions facilitates estimating the isoform expression level. A number of methods have been proposed to deal with the read mapping ambiguity and estimate isoform expression level, such as rSeq\cite{rseq}, RSEM\cite{rsem}, Cufflinks\cite{cufflinks} and BitSeq\cite{bitseq}. Once isoform expression levels are obtained, accurately detecting DE isoforms becomes in need. Some methods have been developed to simultaneously detect DE genes and isoforms and typically are implemented in two steps. The first step is to adopt expression estimation methods to obtain gene and isoform expression. Then in the second step, the obtained expression is used to detect differential expression. We name these methods as two-step methods. For example, EBSeq suggests applying RSEM to quantify expression level and then using an empirical Bayes model for identifying DE genes and isoforms\cite{ebseq}. CuffDiff uses the expression levels obtained from Cufflinks and implements a linear statistical model to evaluate the expression changes of genes and isoforms\cite{cuffdiff}. BitSeq contains two serial steps, i.e., the expression estimation and differential expression analysis. The required expression levels for DE detection are calculated by the step of the expression estimation. Among these approaches, some expression estimation methods, such as Cufflinks and BitSeq, can simultaneously provide the expression level and the associated measurement uncertainty, which accounts for the read mapping ambiguity and sequencing biases. But EBSeq only uses expression levels and ignores the associated measurement uncertainties. It has been testified that such measurement uncertainty is important in DE detection and can lead to improved analysis results in microarray analysis\cite{bgx,pplr,ipplr}. However, the measurement uncertainty receives less attention in RNA-seq analysis. Although CuffDiff\cite{cuffdiff} and BitSeq\cite{bitseq} can account for expression measurement uncertainty in DE detection, CuffDiff often finds fewer DE genes and isoforms than comparable methods\cite{ebseq}, and BitSeq is too time-consuming due to the lowly-efficient Markov Chain Monte Carlo calculations.

In our previous work, we have proposed the GamSeq model which is able to estimate the expression level and the associated measurement uncertainty at both gene and isoform levels\cite{gamseq}. In this contribution, because it has been proven that the expression measurement uncertainty can lead to improved DE analysis, we develop a Bayesian framework, BDSeq, to simultaneously detect DE genes and isoforms with the consideration of the expression measurement uncertainty, which can be obtained from GamSeq. The expression measurement uncertainty can account for both the read mapping ambiguity and sequencing biases. BDSeq adopts two different Bayesian models to integrate the expression measurement uncertainty for DE detection, the basic model and the fast model. Another advantage of BDSeq is that it combines technical or biological replicate measurements when performing DE analysis and can be calculated efficiently. We evaluate BDSeq on three real RAN-seq data sets and compare it with other popular methods. Also, for users' convenience, we develop a GamSeq-BDSeq RNA-seq analysis pipeline. Users can easily apply GamSeq to estimate expression level and then use BDSeq to detect DE genes and isoforms. Meanwhile, the pipeline also provides a user-friendly interface to other approaches.

\section{Methods}
BDSeq simultaneously considers the expression level and the associated measurement uncertainty, which can be obtained from GamSeq. Therefore, before introducing BDSeq, we first provide a brief introduction to GamSeq.

\subsection{Expression level estimation}
GamSeq adopts a Gamma-based model to simulate the generation process of read data. With the adoption of known annotation, GamSeq is able to estimate gene and isoform expression level and the associated measurement uncertainty. This model can deal with the read mapping ambiguity and non-uniform read distribution, which are the major challenging problems for expression estimation.

For each gene, we first design pseudo-probes similar to probes in microarray, and then count the number of reads falling into each pseudo-probe. The parameter $y_{jl}$ represents the read count of $j$-th pseudo-probe in the $l$-th lane. The lane can be deemed as a replicate. We assume
$y_{jl}=\sum_{k}s_{jlk}$, where  $y_{jl}$ is normalized by the sequencing depth of the $l$-th lane and $s_{jlk}$ is the count contributions from the $k$-th isoform. We assume $s_{jlk}$ follows a Gamma distribution, $s_{jlk}\sim Gamma(\alpha_{lk},b_{j})$, where the parameter $\alpha_{lk}$ is a quantity proportional to the abundance of the $k$-th isoform for $l$-th lane and $b_j$ is the sequencing preference of the $j$-th pseudo-probe. The $b_j$ is shared across all lanes and follows a Gamma distribution, $b_{j}\sim Gamma(c,d)$. Under these assumptions, $y_{jl}$  also follows a Gamma distribution,
\begin{equation}
y_{jl} \sim Gamma(\sum_{k}M_{jk}\alpha_{lk},b_{j}),
\end{equation}
where $M_{jk}$ is defined as the indicator function $M_{jk}=1$ if the $j$-th pseudo-probe belongs to the $k$-th isoform, otherwise, $M_{jk}=0$.

The log-likelihood of the observed read counts for a specific gene is
\begin{equation}
\begin{split}
\mathcal {L}&(\{y_{jl}\}|\{\alpha_{lk}\},c,d) = \ln\prod_{j}\prod_{l}P(y_{jl})\\
&=\sum_{j}\ln\int db_{j}P(b_{j}|c,d)\prod_{l}P(y_{jl}|\sum_k M_{jk}\alpha_{lk},b_j).
\end{split}
\end{equation}

The parameters can be estimated using maximum likelihood estimation. Until all parameters reach stable values, we can use these parameters to infer the expression levels and the associated measurement uncertainties of genes and isoforms. For detailed information of GamSeq, please read the original paper\cite{gamseq}.

\subsection{Differentially expression analysis}
\subsubsection{The basic model}

We consider a full Bayesian method for combination of replicated expression. For each gene, a Gaussian distribution of logged expression level across replicates is assumed by
\begin{equation}
\hat{x}_{jl} \sim \mathcal {N}(\mu_j,\lambda^{-1}+\sigma^2_{ij}),
\end{equation}
where $i$ is the index of replicate and $j$ is index of condition. The parameter $\mu_j$ is the mean logged expression level under condition $j$ and $\lambda$ is the inverse of the between-replicate variance. We take the expression measurement uncertainty $\sigma_{ij}^2$ into consideration, which is can be obtained from GamSeq. We make a prior assumption that $\mu_j$ and $\lambda^{-1}$ are independent and assume $\mu_j$ follows a Gaussian prior, $\mu_j\sim \mathcal {N}(\mu_0,\delta_0^{-1})$, where $\mu_0$ and $\delta_0$ are hyper-parameters, on which we adopt noninformative hyperpriors. We assume $\lambda$ follows a conjugate Gamma prior, $\lambda \sim Ga(\alpha,\beta)$.

\begin{figure}[!ht]
  \centering
  \includegraphics[width=0.6\textwidth]{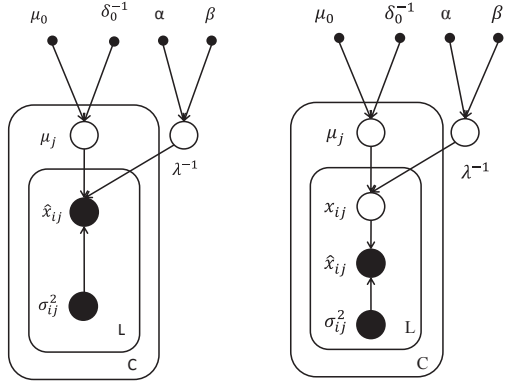}\\
  \caption{The probabilistic graphical models of BDSeqB(left) and BDSeqF(right). The black solid circles represent the observed data and the blank circles represent the hidden parameters. The solid dots represent the hyper-parameters. $C$ is the number of conditions and $L$ is the number of the replicates under one condition.}\label{fig1}
\end{figure}

The basic model is abbreviated to BDSeqB and the probabilistic graphical model is shown in the left subplot in Fig.\ref{fig1}. BDSeqB includes the latent parameters $h=\{\{\mu_j\},\lambda\}$ and the hyper-parameters $\theta=\{\mu_0,\delta_0,\alpha,\beta\}$.

The log-likelihood $\mathcal{L}(\theta,h)$ of the observed data $D$, can be written by the following function,
\begin{equation}
\label{equ1}
\mathcal{L}(\theta,h)=\log P(D|h)=\log\int dhP(D|\theta,h)P(h|\theta).
\end{equation}
However, this integral is intractable. We use a distribution $Q(h)$ over $h$ and Jensen's inequality to get a lower bound of $\mathcal{L}(\theta,h)$. We use the EM algorithm combined with a variational method to optimize the lower bound of $\mathcal{L}(\theta,h)$ and work out the hyper-parameters in M-step\cite{variational}. We assume that the $\mu_j$ and $\lambda$ are independent and obtain the optimization of $Q(\mu_j)$ and $Q(\lambda)$ in E-step:
\begin{equation}
\label{equ2}
Q(\mu_j)\propto \mathcal {N}(\mu_j;\frac{\sum_i\langle s_{ij}\rangle\hat{x}_{ij}+\mu^t_{0}\delta^t_{0}}{\sum_i\langle s_{ij}\rangle+\delta^t_{0}},(\sum_i\langle s_{ij}\rangle+\delta^t_{0})^{-1}),
\end{equation}
\begin{equation}
Q(\mu_j)\propto \mathcal {N}Ga(\lambda;\alpha^t,\beta^t)\prod_{ij}Ga(s_{ij};\frac{3}{2},\frac{1}{2}\langle(\hat{x}_{ij}-\mu_j)^2\rangle),
\end{equation}
where $s_{ij}=(\lambda^{-1}+\sigma^2_{ij})^{-1}$ and $\langle\ast\rangle$ represents the expectation of a function with respect to $Q(\mu_j)$ or $Q(\lambda)$. Since $Q(\lambda)$ is not a standard distribution, the expectation of $Q(\lambda)$ cannot be directly obtained. Importance sampling is thus used to calculate the expectation of $Q(\lambda)$. When the whole EM algorithm is converged, $Q(\mu_j)$ we need in Eq.(\ref{equ2}) is the approximated posterior distribution of mean expression level of condition $j$.
%
%
%\begin{figure}
%  \centering
%  \subfloat[]{
%  \includegraphics[scale=0.35]{BDSeqB}\label{fig1:a}}
%    \subfloat[]{
%  \includegraphics[scale=0.35]{BDSeqF}\label{fig1:b}}
%  \caption{The probabilistic graphical models of BDSeqB(left) and BDSeqF(right). The black solid circles represent the observed data and the blank circles represent the hidden parameters. The solid dots represent the hyper-parameters. $C$ is the number of conditions and $L$ is the number of the replicates under one condition.}\label{fig1}
%\end{figure}

\subsubsection{The fast model}
In the Basic model (BDSeqB), we use important sampling procedure to calculate the expectation of $Q(\lambda)$. As the number of replicates increases, the distribution of $Q(\lambda)$ become more and more flat. Important sampling procedure needs more samples to approximate the expectation of $Q(\lambda)$. This results in low computing efficiency.

In order to avoid the important sampling procedure, we modify the BDSeqB and add a new hidden variable $x_{ij}$, which represents the true expression for each gene on each replicates\cite{fastmodel}. The new model is called BDSeqF, and the probabilistic graphical model is shown in the right subplot in Fig.\ref{fig1}. We assume the hidden variable $x_{ij}$ is Gaussian distributed,
\begin{equation}
x_{ij} \sim \mathcal {N}(\mu_j,\lambda^{-1}),
\end{equation}
where $\mu_j$ and $\lambda^{(-1)}$ has the same prior assumptions as in BDSeqB. Therefore, the observed expression level can be expressed as:
\begin{equation}
\hat{x}_{ij} \sim \mathcal {N}(x_{ij},\sigma^{2}_{ij}),
\end{equation}
The new hierarchical model includes the latent parameters $h=\{\{\mu_j\},\{x_{ij}\},\lambda\}$ and the hyper-parameters $\theta=\{\mu_0,\delta_0,\alpha,\beta\}$.

Similar to BDSeqB, EM algorithm combined with a variational method is used to optimize the lower bound of Eq.(\ref{equ1}). The parameters are assumed to be independent of each other. At E-step, the distributions of $Q(x_{ij})$, $Q(\mu_j)$ and $Q(\lambda)$ are all standard distributions as,
\begin{equation}
Q(x_{ij})\propto \mathcal {N}(x_{ij};\frac{\hat{x}_{ij}+\sigma_{ij}^2\langle\mu_j\rangle\langle\lambda\rangle}{1+\sigma_{ij}^2\langle\lambda\rangle},
\frac{\sigma_{ij}^2}{1+\sigma_{ij}^2\langle\lambda\rangle}),
\end{equation}
\begin{equation}
Q(\mu_{j})\propto \mathcal {N}(\mu_j;\frac{\langle\lambda\rangle\sum_{i}x_{ij}+\mu_0^t\delta_0^t}{\delta_0^t+\sum_i\langle\lambda\rangle},
(\delta_0^t+\sum_i\langle\lambda\rangle)^{-1}),
\end{equation}
\begin{equation}
Q(\lambda)\propto Ga(\lambda;\alpha^t+\sum_{ij}\frac{1}{2},\beta^t+\sum_{ij}\frac{1}{2}\langle(\hat{x}_{ij}-\mu_j)^2\rangle).
\end{equation}

Therefore expectation of them can be quickly obtained in the EM algorithm. BDSeqF avoids the important sampling procedure and can improve the computation efficiency.

\subsubsection{Testing for differential expression}
Once the distribution of $Q(\mu_j)$  is obtained, the posterior distribution $P(\mu_j|D,\theta)$ can be calculated and used to test the DE genes or isoforms between any two conditions.  For instance, there are two conditions (indicated by c1 and c2) and the expression c1 is supposed to be greater than expression c2, then we can compute the probability of expression level by
\begin{equation}
P(\mu_{c1}>\mu_{c2}|D,\theta)=\int_{0}^{+\infty}d(\mu_{c1}>\mu_{c2})P(\mu_{c1}>\mu_{c2}|D,\theta),
\end{equation}
which we refer to as the Probability of Positive Log-Ratio (PPLR). The obtained PPLR values are judged by a level of confidence, like $\alpha$-level in the conventional statistical test. Subsequently, ordering DE gene or isoforms based on the PPLR values produces a ranking of most probable up-regulated and down-regulated genes or isoforms. The similar test has previously been used for the analysis of RNA-seq\cite{bitseq} and microarray data\cite{pplr,ipplr}.

\subsubsection{The BDSeq framework}
Although the BDSeq framework contains two models: BDSeqB and BDSeqF, the processing flow of BDSeq framework is mainly divided into two procedures:
\begin{itemize}
\item \textbf{Estimation:} Combined with the expression level and the associated measurement uncertainty for each replicate, BDSeq estimates the expression level and  standard deviation of each condition.
\item \textbf{Evaluation:} Using the results of above outputs, BDSeq adopts the PPLR value to test the differential expression.
\end{itemize}

The BDSeq framework is listed in Algorithm 1.

\begin{algorithm}
\footnotesize
\caption{: BDSeq}
\begin{algorithmic}
\REQUIRE  Observed data $D$, Condition $C$.
\STATE \textbf{Estimation:}
\REPEAT
\STATE E-step:
\STATE \hspace{0.5cm}$Q(h)^{t+1}=P(h|\theta^t,D)$.
\STATE \hspace{1.0cm}for BDSeqB: $Q(h)=Q(\bm{\mu})Q(\lambda)$.
\STATE \hspace{1.0cm}for BDSeqF: $Q(h)=Q(\bm{X})Q(\bm{\mu})Q(\lambda)$.
\STATE M-step:
\STATE \hspace{0.5cm}$\theta^{t+1}=\arg\max_{\theta}\int dh Q(h)^{t+1}\log P(D|h,\theta)P(h|\theta)$.
\UNTIL{EM algorithm converges.}
\STATE \textbf{Evaluation:}
\STATE \hspace{0.5cm}Calculate $P(\mu_{c1}>\mu_{c2}|D,\theta)$.
\ENSURE  A PPLR value.
\end{algorithmic}
\end{algorithm}

From the Algorithm 1, the major difference of BDSeqB and BDSeqF is E-step. And this results in a distinct complexity between two models. The detailed discussion about complexity analysis and selection of two models is given in Section 3.6.
In this paper, BDSeq mentioned means that BDSeqB and BDSeqF are simultaneously applied.

The BDSeq framework is implemented as an R package.
The R package and documentation are freely available at the website \url{http://parnec.nuaa.edu.cn/liux/GSBD/GamSeq-BDSeq.html}.

\subsection{The GamSeq-BDSeq analysis pipeline}
\begin{figure}[!ht]
  \centering
  \includegraphics[width=0.6\textwidth]{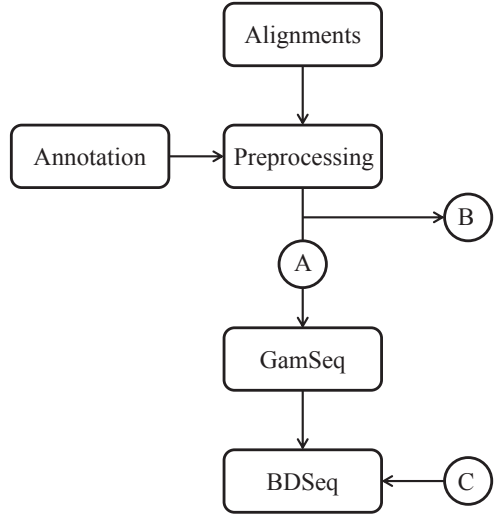}\\
  \caption{The flow diagram of the GamSeq-BDSeq analysis pipeline.}\label{fig2}
\end{figure}

BDSeq requires calculated gene and isoform expression and the associated measurement uncertainty, but is not limited to any particular expression estimation methods. For users' convenience, we develop a GamSeq-BDSeq analysis pipeline, so that users can easily apply GamSeq to estimate expression level and then apply BDSeq to detect DE genes and isoforms. Meanwhile, the pipeline can also provide the read counts of each gene for count-based methods. The flow diagram of GamSeq-BDSeq analysis pipeline is shown in Fig.\ref{fig2}.

In general, RNA-seq analysis pipeline firstly aligns reads to the  reference sequence. Then the alignments are preprocessed according to annotation files to obtain read counts of genes and the mapping relationship between genes and isoforms. The GamSeq-BDSeq analysis pipeline mainly contains the following three routes.

\begin{itemize}
  \item \textbf{Route A}: Users apply GamSeq to estimate the gene and isoform expression level and the associated measurement uncertainties, and then use BDSeq to detect DE genes and isoforms.
      In this paper, BDSeq employs this route to detect differential expression.
  \item \textbf{Route B}: After preprocessing, read counts can be obtained for each gene. Count-based methods can use the counts to detect DE genes. In the following analysis, count-based methods, such as \cite{DESeq,BaySeq,NBPSeq,sSeq,edgeR}, all adopt this route to detect DE genes.
  \item \textbf{Route C}: The input of BDSeq is not specific to expression estimation from any particular expression estimation method. Any expression estimation method, which is able to provide the expression level and the associated measurement uncertainty, can be used by BDSeq to detect DE genes and isoforms.
\end{itemize}

For detailed usage of GamSeq-BDSeq analysis pipeline, please refer to the documentation of the pipeline, which is freely available at our website, \url{http://parnec.nuaa.edu.cn/liux/GSBD/GamSeq-BDSeq.html}.

%In this paper, BDSeq adopts the Route A to detect DE genes and isoforms.

\section{Results and Discussion}
\subsection{Data sets}
In order to evaluate the performance of BDSeq, we use three real RNA-seq data sets. Two are used to detect DE genes, and one is utilized to detect DE isoforms.

We select a real RNA-seq data set from Microarray Quality Control (MAQC) project to evaluate BDSeq for detecting DE genes. The MAQC project compared the multiple whole-genome gene expression profile across various commercial platforms at an unprecedented scale, and is wildly used in evaluating platform performance and testing for various data processing approaches\cite{maqc}. The selected data set (SRA010153) contains two conditions, one from brain tissue (HBR) and the other from a mixture of brain tissue type (UHR). Each condition has seven lanes which can be deemed as seven technical replicates. In the MAQC project, about 1000 genes are validated by the TaqMan qRT-PCR experiment and can be served as ground truth to evaluate the DE analysis methods. We adopt the same strategy in \cite{maqcfc} to filter out 305 genes including 217 DE genes and 88 non-DE genes with high confidence according to qRT-PCR measurements. These 305 genes are used as a "gold standard" to compare methods.

A real human colorectal cancer data set (Griffith) is also used to further evaluate BDSeq for detecting DE genes. The Griffith data set compared fluorouracil (5-FU)-resistant human colorectal cancer cell lines MIP101 against their non-resistant counterpart MIP/5-FU24\cite{griffith}. Each condition has seven lanes which can be taken as technical replicates. There are 192 genes which are assayed by qRT-PCR. We used 20 DE genes and 14 non-DE genes with high confidence as the "gold standard" to compare different DE methods.

A real human breast cancer data set (HBC) is used to evaluate BDSeq for detecting DE isoforms\cite{breast}. This data set includes two condition, human breast cancer cell line (MCF-7) and normal cell line (HME), which have seven and four lanes respectively. Eight isoforms in four genes (TRAP1, ZNF580/1, HIST1H2BD and WISP2) have been validated using qRT-PCR experiment on the same cell lines. We compare different DE methods with the log2 fold change results of DE isoforms based on this qRT-PCR validated data.

\subsection{Accounting for expression measurement uncertainty}

One of the major advantages of BDSeq is accounting for expression measurement uncertainty in DE analysis. We make use of MAQC data set to show the usefulness of expression measurement uncertainty in BDSeq.

\begin{figure*}[!ht]
  \centering
  \includegraphics[width=0.9\textwidth]{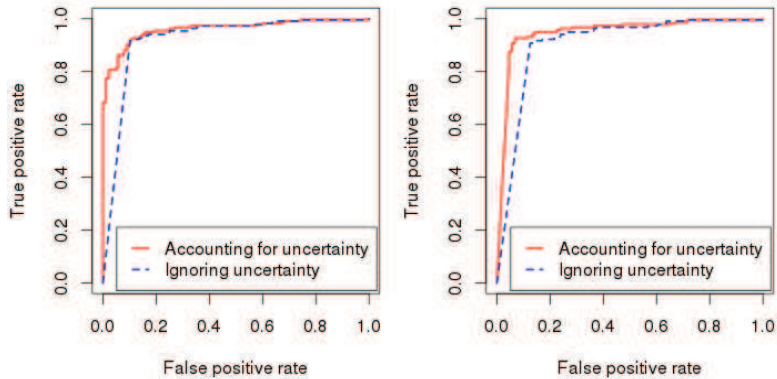}\\
  \caption{Usefulness of expression measurement uncertainty for BDSeq. The left figure is the performance of BDSeqB and the right figure is the performance of BDSeqF. The ROC curves indicate the difference between accounting for and ignoring measurement uncertainty in DE analysis. The solid curve shows the performance of BDSeq accounting for measurement uncertainty, and the dash curve ignores the uncertainty by setting zero measurement uncertainty in BDSeq. }\label{fig3}
\end{figure*}

Fig.\ref{fig3} shows that the expression measurement uncertainty is useful for DE analysis in BDSeq. By considering measurement uncertainty, we obtain better ROC curves for both two models than ignoring measurement uncertainty. The area under ROC curve (AUC) for BDSeqB is 0.9207 if ignoring measurement uncertainty, while 0.9601 if accounting for measurement uncertainty. For BDSeqF, the AUC values for considering and ignoring measurement uncertainty are 0.9472 and 0.9054, respectively. The results demonstrate that accounting for expression measurement uncertainty can significantly improve the accuracy in DE analysis.

\subsection{Detecting DE genes with qRT-PCR validation}

In this paper, we mainly focus on two-step methods which are able to simultaneously detect DE genes and isoforms. Therefore, BDSeq is compared with EBSeq(v.1.1), CuffDiff(v.2.0.2) and BitSeq(v0.4.2). Meanwhile, we also compare BDSeq with two count-based methods, DESeq(v1.8.3) and BaySeq(v1.10.0), to identify DE genes. The input read counts of each gene of the two count-based methods are obtained from Route B in GamSeq-BDSeq analysis pipeline and are normalized by library sizes. EBSeq and CuffDiff use the gene expression levels via RSEM(v.1.2.4) and Cufflinks(v.2.0.2) respectively. BitSeq software includes two stages, the expression estimation and differential expression analysis, and the required gene expression levels are processed through the expression estimation stage. BDSeq makes use of Route A in GamSeq-BDSeq analysis pipeline.

\begin{figure}[!ht]
  \centering
  \includegraphics[width=0.6\textwidth]{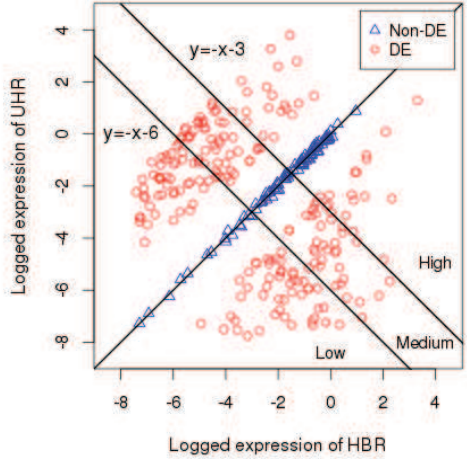}\\
  \caption{The partition of qRT-PCR validated genes in MAQC data set. The scatter plot is drawn using qRT-PCR measurement values of HBR sample against UHR sample. Two lines, $y=-x-6$ and $y=-x-3$, divide the 305 genes into 3 groups, labelled as "low", "medium" and "high".}\label{fig4}
\end{figure}

We use the MAQC data set to evaluate the performance of BDSeq. The MAQC data set contains 305 validated qRT-PCR genes, which are deemed as "gold standard". Fig.\ref{fig4} shows the partition of these qRT-PCR validated genes. 305 genes are divided into three groups, with "low", "medium" and "high" expression respectively, to evaluate the performance of DE methods for genes with different expression levels. The group of all genes is denoted as "all".

%
%\doublerulesep 0.1pt
%\begin{table*}
%\begin{footnotesize}
%\centering
%\caption{Area under ROC curves from different methods for MAQC data set}
%\begin{tabular}{p{2cm}p{1.8cm}p{1.8cm}p{1.8cm}p{1.8cm}p{1.8cm}p{1.8cm}p{1.8cm}}
%\hline\hline\noalign{\smallskip}
% &DESeq	&BaySeq	&BitSeq	&CuffDiff	&EBSeq	&BDSeqB	&BDSeqF\\
%\noalign{\smallskip} \hline
%Low	    &0.9362	&0.9056	&0.8533	&0.8256	&0.9235	&0.8726	&0.8552\\
%Medium	&0.9921	&0.9714	&0.9348	&0.8790	&0.8200	&0.9872	&0.9565\\
%High	&0.9989	&0.9426	&0.8404	&0.9658	&0.8936	&1.0000	&1.0000\\
%All	    &0.9677	&0.9375	&0.8509	&0.8593	&0.8703	&0.9601	&0.9472\\
%  \hline\hline
%\end{tabular}
%\end{footnotesize}
%\footnotesize
%The highest AUC value is highlighted for each group.
%\end{table*}

\begin{figure*}
  \centering
  \includegraphics[width=0.9\textwidth]{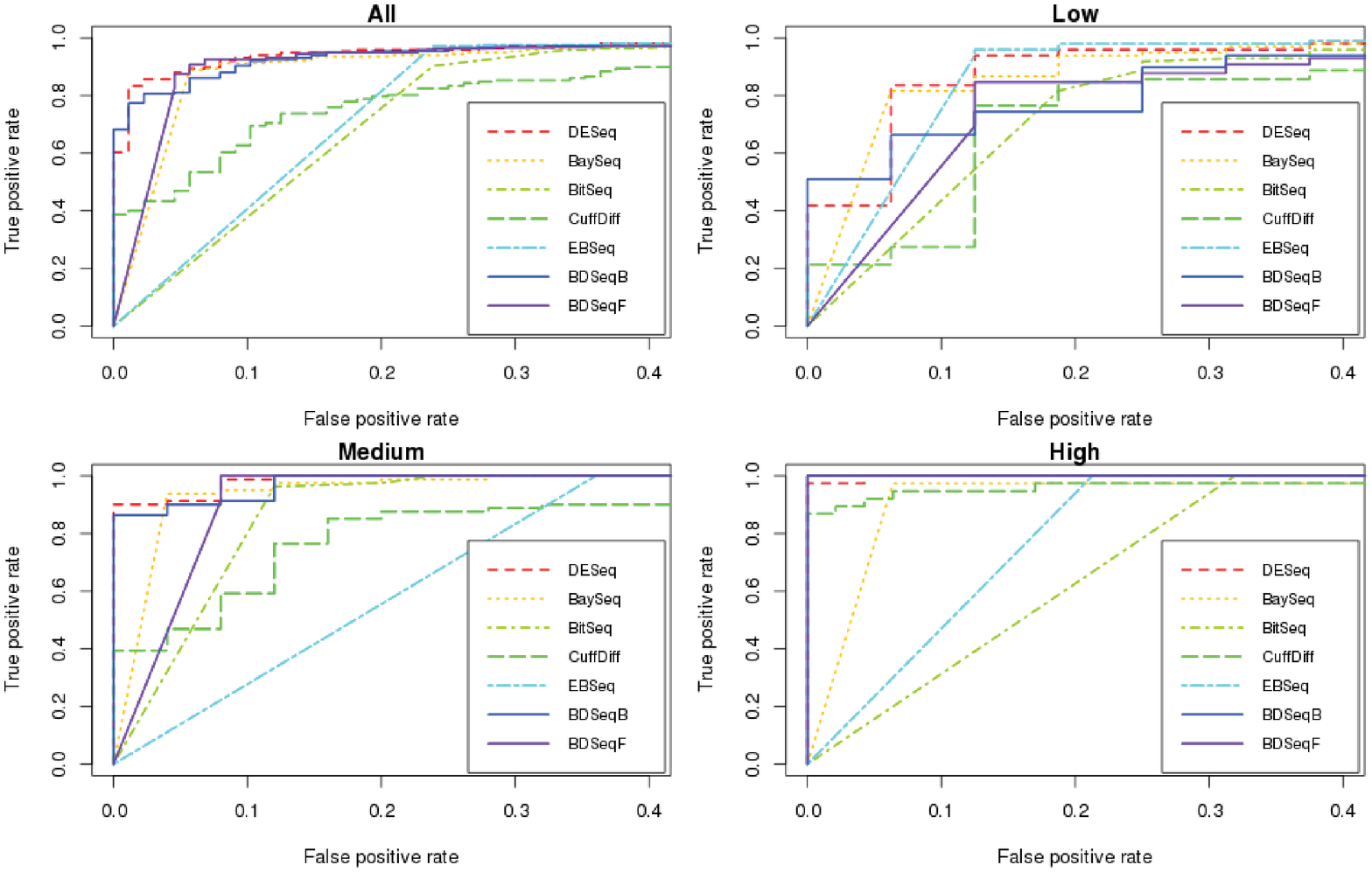}\\
  \caption{ROC curves from different methods for MAQC data set. ROC curves are calculated from eight methods for "all", "low", "medium" and "high" groups. The false positive rate of ROC curves is below 0.4 for focusing on the most significant DE genes.}\label{fig5}
\end{figure*}

For each gene group, we plot ROC curves individually with eight differential expression analysis methods as shown in Fig.\ref{fig5}. The corresponding AUC values are shown in Table \ref{tab1}. In the "all" group, we can see that BDSeqB and BDSeqF obviously outperform the three two-step methods. Compared with count-based methods, BDSeqB and BDSeqF outperform BaySeq, and are slightly worse than DESeq. For the "high" and "medium" groups, results of BDSeqB and BDSeqF are still better than the three two-step methods, and are competitive compared with the count-based methods. However, BDSeqB and BDSeqF both fail in the "low" group. The reason is that there is usually a high level of noise contained in low expression data and most expression estimation methods cannot output accurate expression for low expression genes. In general, BDSeq obtains the most accurate result among the two-step methods for all genes.

\doublerulesep 0.1pt
\begin{table*}[!ht]
\begin{footnotesize}
\centering
\caption{Area under ROC curves from different methods for MAQC and Griffith data set}\label{tab1}
\begin{tabular}{llccccccc}
\hline\hline\noalign{\smallskip}
&  &DESeq	&BaySeq	&BitSeq	&CuffDiff	&EBSeq	&BDSeqB	&BDSeqF\\
\noalign{\smallskip} \hline\noalign{\smallskip}
\multirow{4}{*}{MAQC}&Low	    &\textbf{0.9362}	&0.9056	&0.8533	&0.8256	&0.9235	&0.8726	&0.8552\\
%\cline{2-9}
&Medium	&\textbf{0.9921}	&0.9714	&0.9348	&0.8790	&0.8200	&0.9872	&0.9565\\
%\cline{2-9}
&High	&0.9989	&0.9426	&0.8404	&0.9658	&0.8936	&\textbf{1.0000}	&\textbf{1.0000}\\
%\cline{2-9}
&All	    &\textbf{0.9677}	&0.9375	&0.8509	&0.8593	&0.8703	&0.9601	&0.9472\\
 \noalign{\smallskip} \hline  \hline  \noalign{\smallskip}
Griffith& All	&\textbf{0.8143}	&0.6179	&0.5539	&0.5571&	0.7161	&0.8107	&0.7535\\
\noalign{\smallskip}
\hline\hline
\end{tabular}
\end{footnotesize}
\footnotesize
The highest AUC value is highlighted for each group.
\end{table*}

\begin{figure}[!ht]
  \centering
  \includegraphics[width=0.6\textwidth]{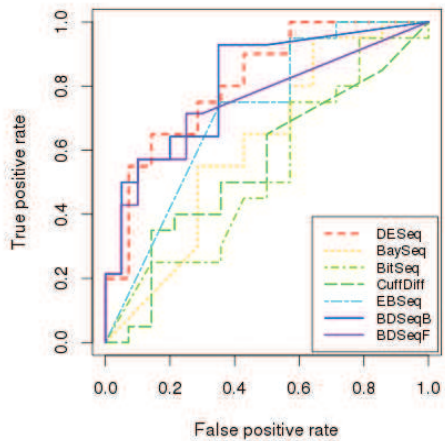}\\
  \caption{ROC curves from different methods for Griffith data set.}\label{fig6}
\end{figure}

Next, we use the Griffith data set to further evaluate the performance of BDSeq. The Griffith data set contains 34 genes validated by qRT-PCR experiments and these 34 genes are deemed as "gold standard" with high confidence. We plot ROC curves for seven methods as shown in Fig.\ref{fig6}, and the corresponding AUC values are shown in Table \ref{tab1}. We can see that DESeq, BDSeqB and BDSeqF obtain the top three results for Griffith data. When false positive rate is below 0.2, BDSeqB and BDSeqF have significantly higher true positive rate than the other five methods, showing the best sensitivity of BDSeq among these competitors.

%
%\doublerulesep 0.1pt
%\begin{table*}
%\begin{footnotesize}
%\centering
%\caption{Area under ROC curves from different methods for Griffith data set}
%\begin{tabular}{p{2cm}p{1.8cm}p{1.8cm}p{1.8cm}p{1.8cm}p{1.8cm}p{1.8cm}p{1.8cm}}
%\hline\hline\noalign{\smallskip}
%& DESeq	&BaySeq	&BitSeq	&CuffDiff	&EBSeq	&BDSeqB	&BDSeqF\\
%\noalign{\smallskip} \hline
%All	&0.8143	&0.6179	&0.5539	&0.5571&	0.7161	&0.8107	&0.7535\\
%  \hline\hline
%\end{tabular}
%\end{footnotesize}
%\footnotesize
%The highest AUC value is highlighted for each group.
%\end{table*}

We have used MAQC and Griffith data sets to evaluate the seven methods in detecting DE genes. From these results, we find that BDSeqB and BDSeqF outperform the two-step methods and also have the competitive performance against the count-based methods. The count-based methods, such as DESeq and BaySeq, generally have better performance than CuffDiff and BitSeq, but have the disability of detecting DE isoforms. In addition to the detection of DE genes, CuffDiff, BitSeq and BDSeq are also able to detect DE isoforms. Therefore, these approaches have broader application in transcriptome analysis.

\subsection{Detecting DE isoforms with qRT-PCR validation}

Because the count-based methods are not able to detect DE isoforms, BDSeq is compared with EBSeq, CuffDiff and BitSeq to detect DE isoforms.

We use the HBC data set to evaluate the performance of BDSeq in detecting DE isoforms. The HBC data set contains two conditions, HME and MCF-7. Eight isoforms are validated by qRT-PCR and considered as "gold standard". From the column of qRT-PCR in Table \ref{tab2}, eight isoforms are all up-regulated DE isoforms, which means that expression levels of these isoforms in MCF-7 cell lines are greater than those in HME cell line. The results of the eight isoforms from various methods are shown in Table \ref{tab2}. We note that all methods except CuffDiff successfully detect seven DE isoforms. For BDSeqB and BDSeqF, the PPLR values of seven consistent isoforms all equal to 1. This means that the seven isoforms are considered as the most significantly DE. Only the results of "uc002cvs.1" isoform are not consistent with qRT-PCR results for all methods. Because the log2 fold change of "uc002cvs.1" obtained from qRT-PCR measurements is 0.5, this isoform is weakly up-regulated, and is thus difficult to be detected. Noteworthily, CuffDiff fails to identify all eight DE isoforms showing the lowest power in DE detection among these methods.

\doublerulesep 0.1pt
\begin{table*}[!ht]
\begin{footnotesize}
\centering

\caption{Results of isoforms between two conditions in HBC data set}\label{tab2}
\begin{tabular}{lcccccc}
\hline\hline\noalign{\smallskip}
   & qRT-PCR&	EBSeq	&CuffDiff&	BitSeq	&BDSeqB	&BDSeqF\\
\noalign{\smallskip} \hline
uc002cvt.2	& DE+	& S(0.0001)	& N(0.9999)	& S(0.9993)	& S(1.0000)	& S(1.0000)\\
uc002cvs.1	& DE+	& N(0.4764)	& N(1.0000)	& N(0.8400)	& N(0.5928)	& N(0.5706)\\
uc002qlq.1	& DE+	& S(0.0001)	& N(0.9999)	& S(0.9994)	& S(1.0000)	& S(1.0000)\\
uc002qlp.1	& DE+	& S(0.0001)	& N(0.9999)	& S(0.9881)	& S(1.0000)	& S(1.0000)\\
uc002xmn.1	& DE+	& S(0.0359)	& N(1.0000)	& S(1.0000)	& S(1.0000)	& S(1.0000)\\
uc002xmo.1	& DE+	& S(0.0000)	& N(0.8511)	& S(1.0000)	& S(1.0000)	& S(1.0000)\\
uc003ngr.1	& DE+	& S(0.0000)	& N(0.2438)	& S(1.0000)	& S(1.0000)	& S(1.0000)\\
uc003ngs.1	& DE+	& S(0.0000)	& N(0.6421)	& S(1.0000)	& S(1.0000)	& S(1.0000)\\
  \hline\hline
\end{tabular}
\end{footnotesize}
\footnotesize
The "DE+" represents that the isoform is up-regulated and expression level of the isoform in MCF-7 cell line is greater than that in HME cell line. "S" represents the isoform is significantly DE isoform and the direction of expression change is consistent with qRT-PCR measurement. "N" represents the isoform is non-DE isoform and the direction of expression change is not consistent with qRT-PCR measurement. The values in brackets for various methods are used to detect DE isoforms. CuffDiff uses false discovery rate (FDR) whereas EBSeq applies posterior probabilities. BitSeq, BDSeqB and BDSeqF all use PPLR values. When the value of CuffDiff or EBSeq is less than 0.05, the corresponding isoform is deemed as significantly DE. In contrast, PPLR value is greater than 0.95, the corresponding isoform is deemed as significantly DE.
\end{table*}

In the real world, due to the lack of isoforms validated by qRT-PCR experiments, we build additional eight comparisons to further evaluate the performance of BDSeq in DE detection. Each comparison consists of two isoforms within a gene under the same condition. We detect differential expression of the two isoforms and compare the regulation relationship of the two isoforms with qRT-PCR results. According to qRT-PCR measurements, eight comparisons are all DE, and contain four up-regulations and four down-regulations. Because the programs of CuffDiff, EBSeq and BitSeq are not able to deal with these types of comparisons, we choose a baseline comparable approach, \emph{t-test}, to detect DE comparisons. Since BitSeq obtains the best results among the above three methods in Table \ref{tab2}, the input values of \emph{t-test} are expression levels estimated by BitSeq. Then we compare BDSeq and \emph{t-test} with qRT-PCR results and show the comparison results in Table \ref{tab3}. We note that BDSeqB and BDSeqF can detect four and five consistent comparisons respectively, but \emph{t-test} detects only two consistent comparisons. BDSeq is obviously better than the baseline comparable approach, \emph{t-test}.

In a word, for detecting DE isoforms between two conditions, BDSeqB and BDSeqF obtain accurate results as well as EBSeq and BitSeq, and are obviously better than CuffDiff. Furthermore, for detecting DE comparisons, our method is also clearly better than the baseline approach, \emph{t-test}.

\doublerulesep 0.1pt
\begin{table*}[!ht]
\begin{footnotesize}
\centering
\caption{Results of comparisons in HBC data set}\label{tab3}
\begin{tabular}{lcccccc}
\hline\hline\noalign{\smallskip}
  Gene	&Comparisons&	Condition	&qRT-PCR	&t-test&	BDSeqB&	BDSeqF\\
\noalign{\smallskip} \hline
\multirow{2}{*}{TRAP1}& \multirow{2}{*}{uc002cvt.2 vs uc002cvs.1} & HME &DE+	&N(0.0000)	&N(0.4927)	&N(0.4940)\\
\cline{3-7}
                                             & & MCF-7&DE+	&N(0.0000)	&N(0.0030)	&N(0.0000)\\
\hline
\multirow{2}{*}{ZNF580/1}& \multirow{2}{*}{uc002qlq.1 vs uc002qlp.1} & HME&DE-	&S(0.0000)	&S(0.0301)	&S(0.0063)\\
\cline{3-7}
                                            &  & MCF-7&DE-	&S(0.0000)	&S(0.0000)	&S(0.0000)\\
\hline
\multirow{2}{*}{WISP2}& \multirow{2}{*}{uc002xmn.1 vs uc002xmo.1} & HME&DE-	&N(0.3991)	&N(0.9997)	&N(1.0000)\\
\cline{3-7}
                                            &  & MCF-7&DE-	&N(0.0801)	&N(0.0790)	&S(0.0410)\\
\hline
\multirow{2}{*}{HIST1H2BD}& \multirow{2}{*}{uc003ngr.1 vs uc003ngs.1} & HME&DE+	&N(0.0025)	&S(1.0000)	&S(1.0000)\\
\cline{3-7}
                                            &  & MCF-7&DE+	&N(0.0002)	&S(1.0000)	&S(1.0000)
\\

                                            \hline\hline
\end{tabular}
\end{footnotesize}
\footnotesize
The "DE+" represents that the expression of second isoform is greater than that of the first isoform within a gene, otherwise, "DE-". "S" represents the comparison is significantly DE and the direction of expression change is consistent with qRT-PCR measurement. "N" represents the comparison is non-DE and the direction of expression change is not consistent with qRT-PCR measurement. The values in brackets for various methods are used to detect DE comparisons. \emph{t-test} uses p-values and BDSeq uses PPLR values. When PPLR value is greater than 0.95, the corresponding comparison is deemed as significantly "DE+". In contrast, PPLR value is less than 0.05, the corresponding comparison is deemed as significantly "DE-". For \emph{t-test}, when p-value is less than 0.05, the corresponding comparison is deemed as significantly "DE+" with the positive log2 fold change of expression levels, whereas the corresponding comparison is deemed as significantly "DE-" with the negative log2 fold change of expression levels.
\end{table*}

\subsection{Detecting DE genes and isoforms without qRT-PCR validation}

We further analysis the Griffith data set to evaluate our method on all genes and isoforms without qRT-PCR validation. After filtering out genes for which total counts over all replicates are less than 10, the data set contains 17041 genes and 101163 isoforms. We use six and three methods to detect DE genes and isoforms, respectively. When PPLR values are less than 0.05 or greater than 0.95, the genes or isoforms are deemed as significantly DE for BitSeq and BDSeq. For other methods all genes or isoforms are found to be significantly DE at a threshold of 0.05.

\begin{figure}[!ht]
  \centering
  \includegraphics[width=0.8\textwidth]{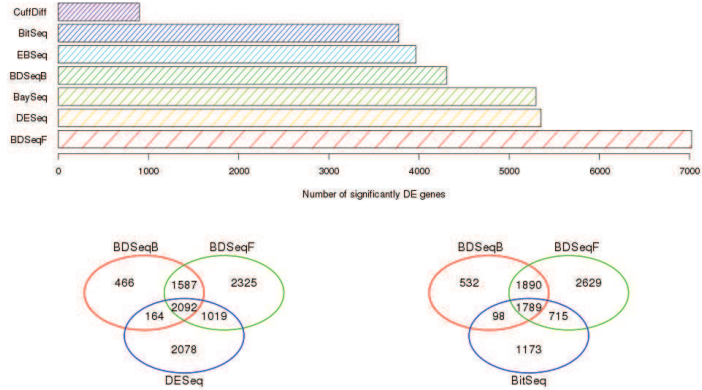}\\
  \caption{Analysis of Griffith data set at the gene level. The panel (top) shows the number of DE genes found between two conditions by various methods. The two Venn diagrams (bottom) show the numbers of DE genes identified by BDSeqB, BDSeqF compared with DESeq and BitSeq, respectively.}\label{fig7}
\end{figure}

First, we analyze the performance of various methods at the gene level. In Fig.\ref{fig7}, the top panel shows the number of genes which are found to be significantly DE between the two conditions. The highest number of DE genes is found by BDSeqF, while CuffDiff returns the least. Next, we study the overlap between the sets of genes called DE by various methods. We select two typical methods, DESeq and BitSeq, to compare with BDSeq. The results are displayed by two Venn diagrams in the lower panel of Fig.\ref{fig7}. From the two Venn diagrams, we note DE genes found by DESeq and BitSeq are to be at a large extent similar to those detected by BDSeqB and BDSeqF. Meanwhile, BDSeqF can detect a fair amount of "unique" DE genes, which are not found by the other methods.

\begin{figure}[!ht]
  \centering
  \includegraphics[width=0.8\textwidth]{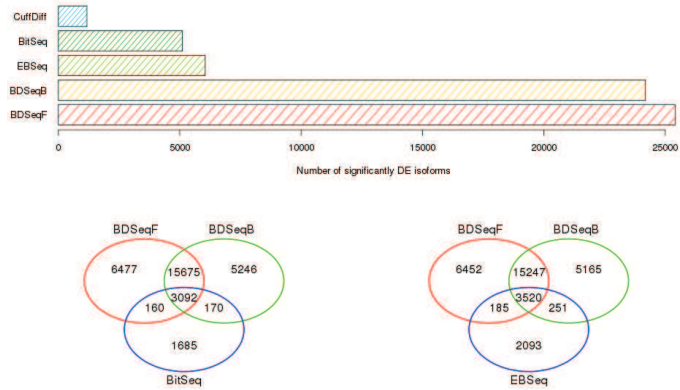}\\
  \caption{Analysis of Griffith data set at the isoform level. The panel (top) shows the number of DE isoforms found between two conditions by various methods. The two Venn diagrams (bottom) show the numbers of DE isoforms identified by BDSeqB, BDSeqF compared with BitSeq and EBSeq, respectively.}\label{fig8}
\end{figure}

We also evaluate the performance of various methods at the isoform level. The top plot in Fig.\ref{fig8} shows the number of DE isoforms which are found between the two conditions. We note that the number of DE isoforms found by BDSeqB and BDSeqF is several times greater than that of EBSeq and BitSeq, while CuffDiff still returns the least number of DE isoforms. In the real world, one or several DE isoforms generally result in the corresponding DE genes between conditions. About 5000 genes are considered as significantly DE from Fig.\ref{fig7} and the average number of isoforms for a gene is 5.94. Hence EBSeq, BitSeq and CuffDiff are too strict to control the number of DE isoforms and lose some DE isoforms, especially CuffDiff. From the two Venn diagrams in Fig.\ref{fig8}, we note that at least half of DE isoforms found by EBSeq and BitSeq are also detected by BDSeqB and BDSeqF.

From the comparisons above, we find that BDSeq has more power to obtain more significantly DE genes and isoforms, especially DE isoforms. Noteworthily, CuffDiff finds the fewest DE genes and isoforms among all methods. This validates the drawback of CuffDiff which often finds fewer DE genes and isoforms as mentioned in Leng et al.\cite{ebseq}.

\subsection{Model selection}

From the above comparisons, BDSeqB and BDSeqF produce the similar performance in detecting DE genes and isoforms. However, due to the adoption of important sampling to approximate the posterior distribution, BDSeqB involves in low efficiency of computation. In order to improve the computational efficiency, BDSeqF are proposed to avoid the inefficient important sampling procedure. We compare BDSeqF with BDSeqB in terms of computational efficiency on Griffith data set. Computation time for the two models is shown in Table \ref{tab4}. We note that BDSeqF is at least four times faster than BDSeqB. As the number of replicates increases, the improvement of computational efficiency for BDSeqF is more obvious compared with BDSeqB.

\doublerulesep 0.1pt
\begin{table}[!ht]
\begin{center}
\begin{footnotesize}
\caption{The computation time (in minutes) of BDSeqB and BDSeqF} \label{tab4}
\begin{tabular}{lccc}
\hline\hline\noalign{\smallskip}
    Replicates No.&Replicates=6& Replicates=10&Replicates=14 \\
%\noalign{\smallskip} \hline\noalign{\smallskip}
%    BDSeqB(gene)& \hspace{0.5cm}53.4 & \hspace{0.5cm}68.3&\hspace{0.5cm}88.5 \\
%    BDSeqF(gene) & \hspace{0.5cm}12.1 & \hspace{0.5cm}12.8 &\hspace{0.5cm}15.3 \\
%    BDSeqB(isoform) & \hspace{0.5cm}469.8 & \hspace{0.5cm}502.8 &\hspace{0.5cm}647.9 \\
%    BDSeqF(isoform) & \hspace{0.5cm}99.3 & \hspace{0.5cm}104.7 &\hspace{0.5cm}116.3 \\\noalign{\smallskip}
%\hline\hline
\noalign{\smallskip} \hline\noalign{\smallskip}
    BDSeqB(gene)& 53.4 &68.3&88.5 \\
    BDSeqF(gene) & 12.1 & 12.8 &15.3 \\
    BDSeqB(isoform) & 469.8 & 502.8 &647.9 \\
    BDSeqF(isoform) & 99.3 & 104.7 &116.3 \\\noalign{\smallskip}
\hline\hline

\end{tabular}
\end{footnotesize}
\end{center}
\footnotesize
Computation time is obtained on a 3.2GHz Quad-Core Intel machine with 16G RAM. After filtering out genes for which total counts over all replicates are less than 10, the Griffith data set contains 17041 genes and 101163 isoforms.

\end{table}

BDSeqB obtains relatively more accurate results than BDSeqF whereas BDSeqF runs more quickly. In practice, the tradeoff between accuracy and computational efficiency is necessary. When users only concern limited number of known genes and isoforms, we recommend using BDSeqB to obtain more accurate results. If users concern the differential expression of genes or isoforms in the whole genome, we recommend choosing BDSeqF for quicker computation and filter out a reduced number of features for further study.

For the workflow of  differential expression analysis, different strategies result in the distinct computational efficiency.
Count-based methods directly utilize the normalized read counts between conditions.
However, two-step methods  require  to calculate the expression level before detecting DE.
The consideration of the various biases and the calculation of isoform expression levels make these methods complex and more time-consuming compared against count-based methods.
In practice, users may choose  appropriate methods according to the purpose of their experiments.
When users only concern  DE genes, we recommend using the count-based methods.
If DE isoforms are of interest, the two-step methods can be a proper solution.

\section{Conclusion}
In this paper, we have proposed a Bayesian framework, BDSeq, which combines technical or biological replicates and considers the expression measurement uncertainty to improve DE analysis. Unlike the popular count-based methods, BDSeq is able to detect not only DE genes but also DE isoforms. It considers the expression measurement uncertainty and makes full use of information in RNA-seq data.
%The expression measurement uncertainty can account for both the read mapping ambiguity and sequencing biases from RNA-seq data.
The results from MAQC data set have  proven that accounting for the expression measurement uncertainty can significantly improve the accuracy of DE analysis. For detecting DE genes, the results from MAQC and Griffith data sets have shown that BDSeq outperforms the two-step methods and also has a competitive performance against the count-based methods. For detecting DE isoforms, BDSeq can obtain competitive results as the two-step methods, EBSeq and BitSeq, and is obviously better than CuffDiff, which is the most popular method for detecting DE isoforms. Meanwhile, we found that BDSeq has more power to find more significantly DE genes and isoforms in the whole genome, especially DE isoforms.
Therefore, BDSeq accounts for expression measurement uncertainty and improves the accuracy of DE analysis.

BDSeq framework adopts two different Bayesian models to integrate the expression measurement uncertainty for DE detection. BDSeqB obtains more accurate results than BDSeqF. However, BDSeqF obviously improves computational efficiency compared to BDSeqB. Therefore, balance between the accuracy and computational efficiency is in need in practice.
%When users only concern limited number of known genes and isoforms, we suggest that users apply BDSeqB to obtain more accurate results. If users want to know differential expression of genes and isoforms in the whole genome, we suggest that users choose BDSeqF for quicker computation and filter out a reduced number of features for further study.

In order to facilitate users, we develop a GamSeq-BDSeq RNA-seq analysis pipeline. In this pipeline, users can easily apply GamSeq to estimate the expression level and the associated measurement uncertainty, and then use BDSeq to detect DE genes and isoforms. Meanwhile, the pipeline also provides a user-friendly interface for other approaches. It can produce the read counts of each gene for count-based methods, and can process expression estimation from other expression estimation methods, which are able to calculate the expression level and the associated measurement uncertainty, for DE analysis.

%Because BDSeq adopts two-step strategy, we have two thoughts to improve performance of detect DE in the future work.

All the above mentioned methods, including BDSeq, only detect differential expression for a single gene. However,
in the real world, biological phenomena usually occur through the interactions of multiple genes via signalling pathways, networks, or other functional relationships. Based on the prior biological knowledge, a number of genes with related functions are grouped together and referred to as a "gene set". Therefore, detecting the differential gene sets using RNA-seq data can provide more useful information to biologists, and is likely lead to more comprehensive biological conclusions.

\end{document}